# Quantum Optical Immunoassay: Upconversion Nanoparticle-based Neutralizing Assay for COVID-19

**Navid Rajil**[1,+,x,†,‡], **Shahriar Esmaeili**[1,+,†,‡], **Benjamin W. Neuman**[1,2,3,+,‡], **Reed Nessler**[1,+,‡], **Hung-Jen Wu**[4,+,‡], **Zhenhuan Yi**[1,+,‡], **Robert W. Brick**[1,+,‡], **Alexei V. Sokolov**[1,5,+,‡], **Philip R. Hemmer**[1,6,7,+,‡], and **Marlan O. Scully**[1,5,*,+,‡]

[1]Institute for Quantum Science and Engineering, Texas A&M university, TX 77843, US
[2]Department of Biology, Texas A&M University, College Station, TX 77843, US
[3]Global Health Research Complex, Texas A&M University, College Station, TX 77843, US
[4]Department of Chemical Engineering, Texas A&M University, College Station, TX 77843, US
[5]Baylor University, Waco, TX 76798, US
[6]Department of Electrical & Computer Engineering, Texas A&M University, College Station, TX 77843, US
[7]Zavoisky Physical-Technical Institute, Federal Research Center "Kazan Scientific Center of RAS", Sibirsky Tract, 420029 Kazan, RU
[*]corresponding.scully@tamu.edu
[+]These authors conceptualized the experiment.
[x]This author built the microscope, prepared samples, acquired data.
[†]These authors processed the data.
[‡]These authors wrote the manuscript.

## ABSTRACT

In a viral pandemic, a few important tests are required for successful containment of the virus and reduction in severity of the infection. Among those tests, a test for the neutralizing ability of an antibody is crucial for assessment of population immunity gained through vaccination, and to test therapeutic value of antibodies made to counter the infections. Here, we report a sensitive technique to detect the relative neutralizing strength of various antibodies against the SARS-CoV-2 virus. We used bright, photostable, background-free, fluorescent upconversion nanoparticles conjugated with SARS-CoV-2 receptor binding domain as a phantom virion. A glass bottom plate coated with angiotensin-converting enzyme 2 (ACE-2) protein imitates the target cells. When no neutralizing IgG antibody was present in the sample, the particles would bind to the ACE-2 with high affinity. In contrast, a neutralizing antibody can prevent particle attachment to the ACE-2-coated substrate. A prototype system consisting of a custom-made confocal microscope was used to quantify particle attachment to the substrate. The sensitivity of this assay can reach 4.0 ng/ml and the dynamic range is from 1.0 ng/ml to 3.2 µg/ml. This is to be compared to 19 ng/ml sensitivity of commercially available kits.

## 1 Introduction

The COVID-19 pandemic has shown how researchers equipped with the proper tools can rapidly translate scientific advances into improvements in healthcare as in the case of rapid viral genome sequencing[1], proliferation of rapid antigen[2,3], antibody[4,5] and nucleic acid tests[6,7], rapid determination of new protein structures[8], and vaccines based on a stabilized version of the viral spike protein[9]. In such a pandemic, vaccination is anticipated to be the main tool to control the rapid spread of infections, and subsequent hospitalizations. Ideally, this paradigm works at its best when every vaccinated individual produces antibodies of sufficient strength and specificity to neutralize the virus. In practice, variations among individuals, population dynamics of antibodies, and frequent mutations of the virus can quickly reduce the effectiveness of vaccines[10–12]. One of the pandemic management tools still lacking improvement is a quick, reliable assay to measure the presence of neutralizing antibodies in serum. This would facilitate decisions on the timing of revaccination, calculations of herd immunity, and provide a probe for ever-growing pool of viral variants[13]. In addition, laboratory-generated antibodies, produced as antibody therapeutic treatments, need to be evaluated with a sensitive test to quantify their neutralizing potential.

The neutralizing antibody is defined by its ability to prevent the virus from interacting with a susceptible cell in a way that leads to infection[14–16]. Neutralizing antibodies first appear about two weeks after vaccination[17], at roughly the same time that protection becomes evident[18]. Current methods to measure neutralizing level of antibodies require live cells and either intact SARS-CoV-2[19] or virus-like particles consisting of a generic shell decorated with SARS-CoV-2 spike proteins[20], and are prohibitive in terms of cost, expertise, and time for wider point of care use.

In the recent years, the fluorescent detection of single biomolecule has gained popularity[21–26]. However, photobleaching of the fluorescent dyes remains challenging for fluorescent readout. In addition, high-resolution microscopy and a longer acquisition time are often needed for single molecule detection, limiting its applications[20–26]. To resolve the photobleaching issue, the use of lanthanide doped upconversion nanoparticles (UCNPs) as fluorescent tags has been proven to be beneficial. For instance, Farka et al. use UCNPs as their fluorescent tags to detect prostate specific antigen with a sensitivity of 1.2 pg/ml (42 fM) in 25% serum[27]. UCNPs can be excited by infrared lasers and remain stable after long exposure. In addition, the use of a high-power laser in a wide-field illumination configuration can increase the fluorescent signals and reduce the acquisition time, leading to faster measurements. It has been shown that using highly stable UCNPs could improve the limit of detection (LOD) of upconversion-linked immunosorbent assay (ULISA) by an order of magnitude compared to commercially available assays[26,27]. More sensitive optical readout can also improve the LOD in such bioassays[28,29]. These progressions toward single molecule detection are pushing the sensitivity, specificity, and LOD beyond what was once theoretically possible[30–32].

Here, we show a proof of concept for a safe, simple, low-cost assay to determine the neutralizing activity of anti-SARS-CoV-2 antibodies using the tools of quantum optics. We use fluorescent UCNPs to measure the relative effectiveness of antibodies in preventing the binding of SARS-CoV-2 receptor-binding domain (RBD) to angiotensin-converting enzyme 2 (ACE-2). The method proposed in this study is in keeping with the principle that SARS-CoV-2 neutralizing antibodies could prevent the interaction between the RBD of the viral spike protein with the ectodomain of ACE-2[33]. For one clone of antibody used, we calculated the midpoint inflection point (IC50) to be 12 ng/ml (80 pM) and the LOD, defined as a concentration two standard deviations lower than the mean negative control value, to be 4 ng/ml (33 pM).

## 2 Results and discussion

### 2.1 The Upconversion nanoparticle-based Neutralizing Immunoassay Kit (UNIK)

The basic principle of the upconversion nanoparticle-based neutralizing immunoassay kit (UNIK) is shown in Figure 1. The assay relies on the natural affinity between RBD and ACE-2 protein. To fully take advantage of this property, we employed streptavidin conjugated upconversion nanoparticles and biotinylated RBD to produce the upconversion nanoparticle phantom virion (UCPV). If there is no antibody present in the sample (or if the antibodies present in the sample are non-neutralizing), the phantom virus particles will bind to the substrate without any obstruction. As a result, images taken from these samples will show high count of particles (Figure 1a and 1b). On the other hand, if the antibody is effectively neutralizing the RBD, then the binding of phantom virus and ACE-2 will be hindered, thus a lower count of particles will be observed in the images (Figure 1c and 1d), compared to the negative control sample with no antibody present, as shown in Figure 1a and Figure 1b.

### 2.2 Assessment of ACE-2/polydopamine coated plates

Glass bottom plates were coated with ACE-2/polydopamine mixtures. The activity of ACE-2 was evaluated by measuring the binding between ACE-2 and SARS-CoV-2 RBD that is linked to mouse IgG Fc tag. RBD was further detected by goat anti-mouse antibody with Alexa fluor 633. The fluorescent spectra are shown in figure S4 (b, d, e, h). To make sure that non-specific bindings or autofluorescence signals were minimal, several control experiments were performed. (Figure S4 a-h). Figure S4a shows the positive control test, in which the plates were coated with ACE-2/polydopamine mixture and then blocked with 5% BSA solution (supplementary materials). The next layer was RBD with mouse Fc tag which was detected with goat anti-mouse antibody with Alexa fluor 633. Figure S4b shows the spectrum of the Alexa fluor obtained with a 638 nm laser. This spectrum clearly shows the positivity of this sample. Figure S4c shows the negative control test, in which instead of RBD with Fc tag, the plate was incubated 1×PBS as a negative control sample. Figure S4d shows the spectrum of the Alexa fluor obtained with 638 nm laser from this sample. This spectrum clearly shows the negativity of this sample, as there is only background readout signal. Figure S4e shows the control sample which is missing the goat anti-mouse antibody with Alexa fluor 633. Since in these measurements the excitation laser was the 638 nm laser, there was a possibility of auto fluorescent background from any of the elements on the plate. To check if there was any auto fluorescent, we prepared this sample and scanned it. Figure S4f shows the spectrum of the Alexa fluor obtained with the 638 nm laser from sample shown in Figure S4e. This spectrum clearly shows the negativity of this sample, as there is only background readout signal. So, there are minimal auto fluorescent signals from other elements on the plate. Figure S4g shows the control test which is missing the ACE-2 protein. This plate was coated with mixture of 1×PBS and polydopamine and then blocked with 5% BSA. The purpose of this test was to measure the extent of non-specific binding of RBD with mouse Fc tag and secondary Alexa fluor conjugated antibody complexes with ACE-2-coated plates. Figure S4h shows the spectrum of the Alexa fluor obtained with 638 nm laser from figure S4g. This spectrum shows a small background in this sample. However, the positive signal shown in figure S4b is approximately 16 times larger than this background.

## 2.3 Non-specific binding

One of the concerns in using any type of plate for bioassays is non-specific binding. The non-specific binding between the phantom virus and ACE-2 coated plates can increase the background signal, thus decrease the LOD and sensitivity. Polydopamine molecules, which are positively charged, can bind to the phantom virus without involvement of ACE-2 and RBD. In addition, any imperfection on the plate's coating or UCPVs can increase the non-specific binding. To assess this, we prepared ACE-2/polydopamine-coated plates and 1 ×PBS/polydopamine coated plates, both blocked with 5% BSA to show that phantom virus particles only bind to substrate when ACE-2 protein is mixed with polydopamine and plated on the substrate. To test the affinity of the particles with the ACE-2 coated plates and compare it with blank and blocked plate (no antibody was used), we imaged the edge of the coated area (the location of the edge was found by rough marking on the glass plate and the coffee-ring effect of the coated area after it was dried). Figure 2a shows the 10 µl coating of polydopamine/ACE-2 protein. Since only a certain area of each well was coated, we expected to find the particles only in that coated region. Figure 2b shows very high binding between phantom virus particles and the coated area, while the uncoated area to the side did not show any non-specific binding to between UCPV and the blocked blank glass coverslip (effective imaged area is 145 µm by 145 µm).

We also coated plates with a polydopamine and PBS mixture and blocked them with 5% BSA blocking buffer to study nonspecific bindings that may rise between polydopamine and UCPVs (Figure S5). We expected to see a very low count of particles on these plates, based on this assumption that there is no affinity between UCPVs and polydopamine. Fig. S5 shows three images of three different areas of the same sample taken from the center of the coated area, and only a few of particles are visible in the images (effective imaged area is 145 µm by 145 µm). The particles appear as small diffraction-limited green spots on the dark background of the images. These results (Figure 2 and supplementary materials Figure S4 and S5) prove that the binding between ACE-2-coated area and UCPV is specifically caused by the natural affinity between RBD and ACE-2 proteins. It is important to note that those non-specific bindings are due to surface imperfections of the substrate and the particles, as well as the protein coating integrity of the particles and substrate. For instance, excessive sonication (which is a step of the UCPV preparation procedure; see supplementary materials) can damage and denature the protein coating of the nanoparticles, either due to excessive heating or high-pressure waves generated by bath sonicator inside the nanoparticle vials. Optimization of every step and paying attention to such details can decrease the amount of non-specific binding.

## 2.4 IC50 and Hill coefficient

To test the neutralizing ability of the antibodies using UCPVs, we made a serial dilution of the antibody clones NN54, T01KHu, and CR3022 and mixed equal volumes of the antibody dilutions with equal volume and concentrations of UCPVs (Table 1). The dilutions were calculated such that the final sample volume on each plate was the same for all samples as was the concentration of UCPVs. But the concentration of the antibodies was different in each sample (i.e., the ratio of particles to antibody was different for each sample). Table 1 shows the final antibody concentration for each data point.

According to the manufacturer's datasheet for neutralizing antibody NN54, the ELISA-based neutralizing assay kit performed on this antibody showed an average IC50 point (defined below) of 0.857 nM (0.129 µg/ml)[34]. As for neutralizing antibody T01KHu, the manufacturer reports the lowest IC50 point to be at 0.1 µg/ml.[35] For CR3022, it has been reported that this antibody does not block binding of RBD with ACE-2 protein[36].

IC50 point in current work is defined as the concentration where the signal count is (maximum count − minimum count)/2 estimated by the fitting 4-parameter logistic function:

$$Y = A + \frac{B-A}{1+(Conc/IC50)^{hc}}, \tag{1}$$

where Y is the total count, A is the minimum count, B is the maximum count, Conc in the concentration of antibody used, IC50 is the concentration of antibodies at which the count is at 50% and hc is the Hill coefficient (see supplementary materials for fitted functions).

The IC50 points are calculated to be 12 ng/ml (80 pM) and 138 ng/ml (933 pM) for NN54 and T01KHu (Figure 3b), respectively. The assay is also capable of differentiating between antibodies' respective Hill coefficients in the context of their interaction with UCPVs. The Hill coefficient has been used to measure the cooperativity of multivalent binding systems[37,38]. In the dose-response curves, the Hill coefficients for NN54 and T01KHu were calculated to be 1.148 and 4.0836, respectively. Comparing with NN54, the binding of T01KHu is closer to multiple ligand interactions. These parameters were calculated by fitting the 4-parameter logistic function to the data sets using an online tool (supplementary materials).[33] Thus, we can differentiate between antibodies in terms of their strength (IC50) and cooperativity (Hill coefficients) in binding to the UCPVs.

Figure 3a and 3b show the effectiveness of neutralizing antibodies NN54 and T01KHu as their concentration increases. They also show that the non-neutralizing clone CR3022 does not prevent UCPVs from binding to ACE-2 coated plates. Our assay can differentiate the IC50 point with high sensitivity and determine the antibody with higher affinity without the use of enzymatic enhancement in ELISA. This assay also shows differences in the Hill coefficients for these two antibodies, which shows that T01KHu is a multiligand interaction while NN54 seems to be a single ligand interaction.

One should be cautious when comparing the results in Figure 3 with reported results from other sources, such as ones reported by the manufacturers of the antibodies and tests. For instance, manufacturer of NN54 reports two IC50s for the same antibody from two different neutralizing tests. In one, they report an average IC50 of 1.41 µg/ml obtained from neutralizing assays involving 293T/ACE-2 cells[34]. These cells were infected with Pseudotyped Luciferase rSARS-CoV-2 Spike and the concentration of neutralizing antibody was changed to see the how many cells were not infected by the spike[34]. In the other test, they report an IC50 of 0.129 µg/ml measured using an inhibitor screening ELISA kit. The question of what the correct IC50 value is, seems irrelevant since the parameters of these tests are different, so as their goals. In short, each test is optimized for certain dynamic range and specific LOD.

To understand how test parameters interplay with LOD for instance, one can take note of the ratio of protein–ligand complexes and total protein molecules ($\theta$ value). One can assume, for simplicity, that ACE-2 is the protein and UCPV is the ligand in our test. The combination of enzymatic reaction and RBD in a neutralizing ELISA test is equivalent to the UCPV in our test. The ratio of protein-ligand complexes to the total proteins, $\theta$, is (see supplementary materials for proof):

$$\theta = \frac{[PL]}{[P]_t} = \frac{([P]_t + [L]_t + [K]_d) - \sqrt{([P]_t + [L]_t + [K]_d)^2 - 4[P]_t[L]_t}}{2[P]_t}, \quad (2)$$

where $\theta$ is the ratio of protein molecules bound to ligand. [PL], $[P]_t$, $[L]_t$, and $K_d$ are total concentration of protein ligand complex, total concentration of protein, total concentration of ligand, and the dissociation constant of protein and ligand respectively. The concentration of $[L]_t$ when $\theta = 0.5$ (IC50 concentration) can be derived from equation 2 with simple algebra as

$$[L]_t|_{IC50} = \frac{1}{2}[P]_t + K_d. \quad (3)$$

As can be seen, the IC50 concentration in reality depends on two parameters. One is the total protein concentration and the other is the $K_d$ value. Using equation 3, for the more complex case of our assay, we can derive the following relation for the antibody concentration (see supplementary materials for proof)

$$[Ab]|_{IC50} = [UCPV]_{total} - K_d^{(2)} + \frac{2[UCPV]_{total}K_d^{(2)}}{[ACE] + 2K_d^{(1)}} - \frac{1}{2}[ACE] - K_d^{(1)} \quad (4)$$

where $[UCPV]_{total}$ is the total UCNP concentration, [ACE] is the total ACE-2 protein concentration, $K_d^{(1)}$ is the dissociation constant between UCPV and ACE-2, and $K_d^{(2)}$ is the dissociation constant between UCPV and neutralizing antibody. In our work, only the UCPV concentration and ACE-2 concentration can be controlled and manipulated to reduce the IC50 concentration. Supplementary figure S8 shows the changes in $[AB]|_{Ic50}$ as a function of ACE-2 concentration, for different values of UCPV concentrations. As can be seen, we needed to maximize the amount of ACE-2 protein on the substrate, while optimizing the UCPV concentration to the lowest amount possible. Rationally, by decreasing UCPVs, we reduced the number of antibodies needed to fully block them, while by maximizing ACE-2 protein we increased the number of unblocked UCPVs captured on the substrate.

In the case of a test such as ELISA, reducing the RBD concentration means fewer actual RBD–ACE complexes will be available to be detected later on (through anti-ligand secondary antibodies and enzymatic enhancement, fluorescent dyes, etc). Since the LOD is an arbitrary choice, and we can choose IC50 concentration for this simple examples, we can conclude that different tests involve different amounts of protein concentrations and are optimized for specific dynamic ranges and different LOD.

A keen observer may ask: for NN54 and T01KHu antibodies, why do manufacturers report similar IC50s concentrations of 0.129 µg/ml and 0.1 µg/ml respectively. Figure S9C illustrates the reason for similar results from companies. When we set $[P]_t = 19$ nM; $K_d = 1$ nM and $[P]_t = 1$ nM, $K_d = 10$ nM, we see that both cases have the same IC50 concentrations where clearly, we assumed different $K_d$ values. The difference in the conditions of the manufacturer's test is perhaps the reason for their similar results.

The advantage of UNIK is apparent from two important factors. First, it can differentiate between IC50 concentrations and Hill coefficients of two different neutralizing antibodies. Second, although the limit of detection (LOD, defined Section 2.7) is a function of both UCPV concentration and the affinity between antibody and UCPV, and affinity between UCPV and ACE-2 (the corresponding $K_d$ values, as shown in equation 4), its LOD is an order of magnitude better that of cited commercial tests, among which the best LOD is reported to be 19 ng/ml[39]. Our results shows that proper optimization of UCPV's concentration while maximizing the number of RBD per UCNP (1200:1) and maximizing ACE-2 protein on the substrate can improve LOD. This is because we can detect single molecule bindings and as such we can reduce UCPV concentration to such low amounts that a lower concentration of neutralizing antibody will be needed to block them while UCPVs can still be detected (described in equation 4). The dependence of LOD to $K_d$ value was also shown by S. Zhang et al[32]. As depicted by equation 4, factors that play a role in UNIK's LOD are $K_d$ value between UCPV and ACE-2 protein, $K_d$ value between the UCPV and antibody, total concentration of UCPV, and total concentration of ACE-2 protein. It is best to derive such equation for every specific assay to maximize the improvement in the LOD.

### 2.5 UCPVs concentration optimization

In neutralizing assays such as UNIK, the antigen concentration plays an important role. In this work, RBD is the antigen and it is pre-bound to the UCNP. The concentration of RBD in the assay depends on two factors, number of RBD bound to each UCNP and final working concentration of RBD. Accordingly, to control the concentration of RBD in the assay there are two methods one could use. It is possible to keep the working concentration of UCPVs constant and optimize the number of RBD per UCNPs[40] or keep the ratio of RBD to UCNPs constant and optimize the concentration of UCPVs (as it has been done in[27]). We decided to choose the latter for the reason that it is necessary to optimize the concentration of UCPVs for a given RBD to UCNP ratio i.e. for any given ratio, one needs to choose the concentration of UCPVs in the linear region with the highest slope (0.1 µg/ml and 1 µg/ml in supplementary Figure S3) to achieve the highest sensitivity. As such, we maximized the amount of RBD per UCNP (1200:1), and optimized the concentration of final UCPV. Figure S3 shows the result of concentration optimization for our experiment. In the region between 0.1 µg/ml and 1 µg/ml the average number of particles counted per image changes rapidly, increasing with concentration of UCPVs and in this region, for the ratio of RBD to UCNP that we used, we have the maximum sensitivity. A slight decrease in the UCPV concentration due to presence of neutralizing antibodies will cause measurable changes in the countable particles in the images.

### 2.6 Non-neutralizing antibody

It is important to differentiate an antibody that binds but does not neutralize infectivity from a truly neutralizing antibody, which would provide direct protection against infections. In the case of clone CR3022, its binding does not block the binding site on the RBD specific for the ACE-2 protein[36]. In a separate experiment, serial dilutions of CR3022 IgG antibody were mixed with UCPVs and tested (see descriptions in the supplementary materials). The results, illustrated in Figure 3a, show that the average particle counts per scan in all these dilutions stay relatively close to the control sample (no statistical significance, $p$-values > 0.1, see supplementary materials). In this work, we define neutralizing activity as binding of the antibody to RBD at the location where ACE-2 protein would bind. As such, CR3022 is not a neutralizing antibody[36].

### 2.7 Detection limit

The limit of detection (LOD) for this assay is defined as the concentration with a count of two times of negative control's standard deviation below negative control's average count[39]. Using this definition and the calibration curves, we estimated the LOD of this assay for both neutralizing antibodies, see supplementary materials for details. The LOD for NN54 and T01KHu are estimated to be 0.004 µg/ml and 0.128 µg/ml respectively. In addition, based on the $p$-values of each data point, we can conclude that there is a statistical significance between 0.00323 µg/ml ($p$-value 0.19) and 0.0323 µg/ml ($p$-value 0.0034). Thus, in practice the detection limit for NN54 can be assumed to be 0.0323 µg/ml. For the case of T01KHu, the statistical significance is first observed between 0.0968 µg/ml ($p$-value 0.71) and 0.196 µg/ml ($p$-value 0.0065) and as a result, the detection limit for this antibody can be assumed to be 0.196 µg/ml.

### 2.8 Assay modifiability

Other variations of this assay are also possible, adjusted for SARS-CoV-2 variants or other viral species. To modify this assay, one can place the RBD of other SARS-CoV-2 variants on the particles and ACE-2 protein on the substrate. It is also possible to use multiple upconversion nanoparticle types with different fluorescent emissions for each RBD variant. For instance, we can conjugate $NaYF_4$:Yb/Tm particles (excitation peak at 980 nm, emission bands around 375 nm and 450 nm) with the Alpha variant of SARS-CoV-2 RBD, and $NaYF_4$:Yb/Er particles (excitation peak at 980 nm, emission bands around 550 nm and 650 nm) with the Delta variant of SARS-CoV-2 RBD. Thus, we can test the antibodies against both variants of virus simultaneously. This is part of our future study plan.

## 3 Conclusion

We have demonstrated that UNIK can be used effectively for determination of neutralizing activity of COVID-19 antibodies. We show that with proper optimization, we can detect the antibody for SARS-CoV-2 virus. Although the limit of detection is dependent on the concentration of RBD as well as the affinity of antibody and RBD, as seen in the case of NN54 and T01KHu antibodies, we report that the lowest detection limit for this assay was 4 ng/ml (27 pM, calculated for neutralizing antibody clone NN54). A paper-based ELISA test for detection of COVID-19 antibodies reported 9.00 ng/µl (i.e., 9.00 µg/ml) limit of detection[41]. A readily available commercial neutralizing assay from Cayman Chemical reports a LOD of 19 ng/ml[39]. The assessment of the performance of the assay with blood serum samples as well as measuring the receiver operating characteristic curve (ROC curve) using human convalescent blood plasma, as well as other variation of the assay mentioned before are subjects of our future studies. We also showed that one must be cautious when defining the LOD, since both measured parameters depend on various factors, resulting in various LOD even under the same conditions but with different samples.

## 4 Methods

All the incubation steps in this section were performed at room temperature unless mentioned otherwise.

Upconversion nanoparticle phantom virions (UCPV) were prepared as follows. Briefly, the streptavidin coated upconvertion nanoparticles were purchased from Creative Diagnostics (part numbers of all materials are listed in supplementary materials Table S1). The they were diluted to 200 µl and concentration of 0.5 mg/ml and sonicated for 10 minutes. Then 10 µl of 0.2 mg/ml biotinylated RBD was added to it and left on the vortex mixer for 1 hour at lowest speed. Subsequently, particles were washed 3 times by centrifuging, replacing the supernatant with fresh assay buffer, and sonicating for 10 minutes (more specific details on sonication and particle wash are in the supplementary materials). Then, the UCPVs where diluted down to 0.4 µg=ml at a volume of 4 ml, sterile filtered using 0.2 µm cellulose acetate syringe filters, and kept in 4 C until use. There are more details on how the concentration of UCVPs was selected in supplementary materials.

The optimization of UCPVs concentration was as follows. We prepared 4 different concentrations of UCPVs (0.1 µg/ml, 0.4 µg/ml, 1 µg/ml, 10 µg/ml) and plated them on on the blocked plates as described below. No antibody was used in this measurement. Subsequently we took 5 images of each concentration, counted the particles and averaged the number per image. The results are shown in figure S3.

To prepare the Nunc Lab-Tek II coverglass plates, we mixed 0.75 mg/ml ACE-2 protein with 2 mg/ml polydopamine solution at 1:1 ratio (more details in the supplementary materials) and plated the solution on the coverglass plate wells. The plates were incubated for 2 hours and kept inside a humidity chamber to prevent drying. The plates were then washed 4 times with assay buffer (1× PBS, 0.5% BSA, 0.1% tween-20) and blocked with 5% BSA solution (1×PBS, 5% BSA, 0.1% tween-20) for 1 hour. Then the plates where washed again 4 times with assay buffer and used immediately. There are more details on validation of plates explained in the Supplementary materials.

To perform the upconversion nanoparticle-based neutralizing immunoassay, different concentrations of the antibodies were prepared (Table 1). Then, 10 µl of each concentration was added to a separate 300 µl of 0.4 µg/ml UCVP solution and left on the mixer for 1 hour. Then, each of UCPV and antibody mixes was added to a separate wells of a prepared Nunc Lab-Tek II 8-well plate and incubated for 1 hour. After incubation, the plates were washed 4 times with assay buffer and kept in 4 C until measurements. This procedure was repeated 3 times for the 3 different antibodies.

To count the number of nanoparticles on the plates, 10 images of each well were obtained using a custom-made confocal microscope (details of the system in supplementary materials). Then, the particles, observed as bright spots in the images (Figure 1b and Figure 1d), were counted and recorded for each final concentration of each antibody (Table 1) using a custom made program in Mathematica software. The counts of 10 images of each data point were averaged for the 3 repetitions for that antibody. More details on how the images were taken and processed can be found in supplementary materials.

## Acknowledgement

The authors thank Jane Pryor, Arash Azizi, and Sahar Delfan for helpful discussions. S.E. was supported by the Herman F. Heep and Minnie Belle Heep Texas A&M University Endowed Fund held/administered by the Texas A&M Foundation.

This research was supported by Air Force Office of Scientific Research (Award No. FA9550-20-1-0366 DEF), Office of Naval Research (Award No. N00014-20-1-2184), Robert A. Welch Foundation (Grants No. A-1261, A-1547), National Science Foundation (Grant No. PHY-2013771, PHY-1820930, and ECCS-2032589), National Institutes of Health (Award No. R03AI139650 and R21AI149383), and King Abdulaziz City for Science and Technology (KACST).

## Author contributions statement

N.R., S.E., B.W.N., H.J.W., Z.Y., R.W.B., A.V.S., P.R.H., and M.O.S. conceptualized the experiment. N.R. built the microscope, prepared samples, acquired data, and processed the data. S.E. processed the data. N.R., S.E., B.W.N, R.N., H.J.W, Z.Y., R.W.B, A.V.S, P.R.H, and M.O.S wrote the manuscript.

## Additional information

### Competing Interests statement

The authors declare no competing interests.

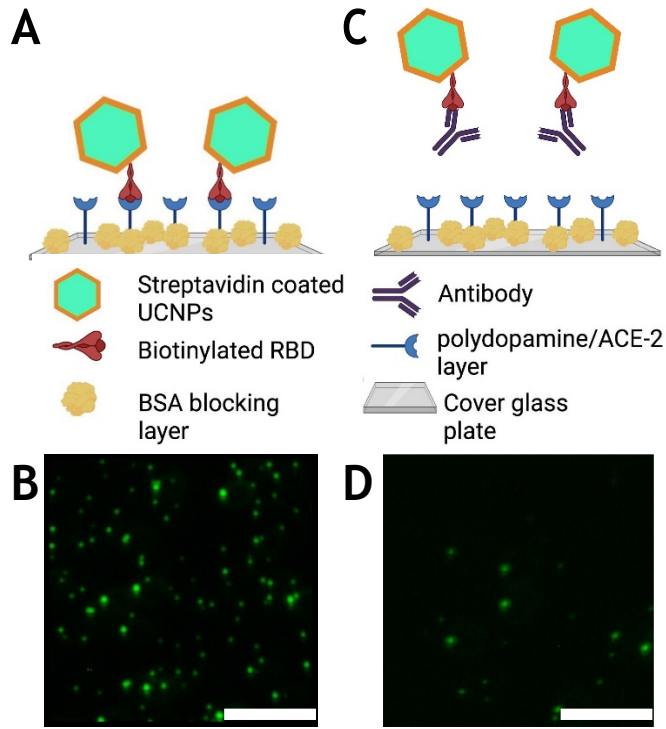

**Figure 1.** Schematic illustration of the upconversion-based neutralizing assay. A, B) When the antibody is not present (or it is not neutralizing), the phantom virion complex will bind to the ACE-2-coated substrate and particles can be imaged and counted as shown in (B). The concentration of UCPVs was 0.4 μg/ml and no antibodies were present in the solution. C, D) When the antibody is present and it is neutralizing, it will prevent the phantom virus complex from binding to ACE-2-coated substrate and as a result, fewer fluorescent particles will be observed compared to the negative control as shown in (d). The concentration of UCPVs was 0.4 μg/ml and the concentration of the antibody was 3.23 μg/ml. Scale bars represent 15 μm.

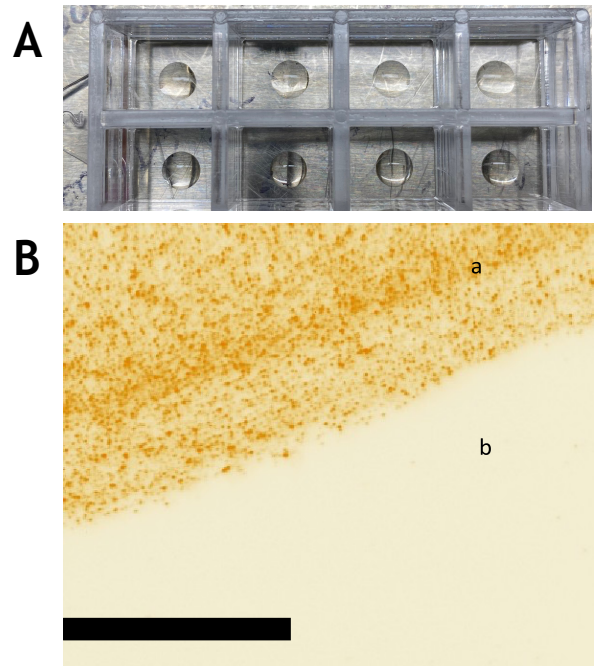

**Figure 2.** Affinity of UCPV and ACE-2 coated area. A) Typical configuration of the coated area on the Nunc LabTek II 8-well dishes with bottom cover glass. The volume of the ACE-2 coating was (10 µl). For the rest of the steps in all experiments, the whole well was filled (as described in methods and supplementary materials). B) After polydopamine/ACE-2 coating and blocking, we incubated the plate with UCPV solution (10 µg/ml). The image was taken from the edge of the coated area. The area coated with ACE-2 (a in Fig. 2B) shows a high fluorescence particle count, while the uncoated and BSA blocked area shows no particles at all (b in Fig 2B). This shows a relatively very minimal nonspecific binding between the blocked cover glass and UCPVs.

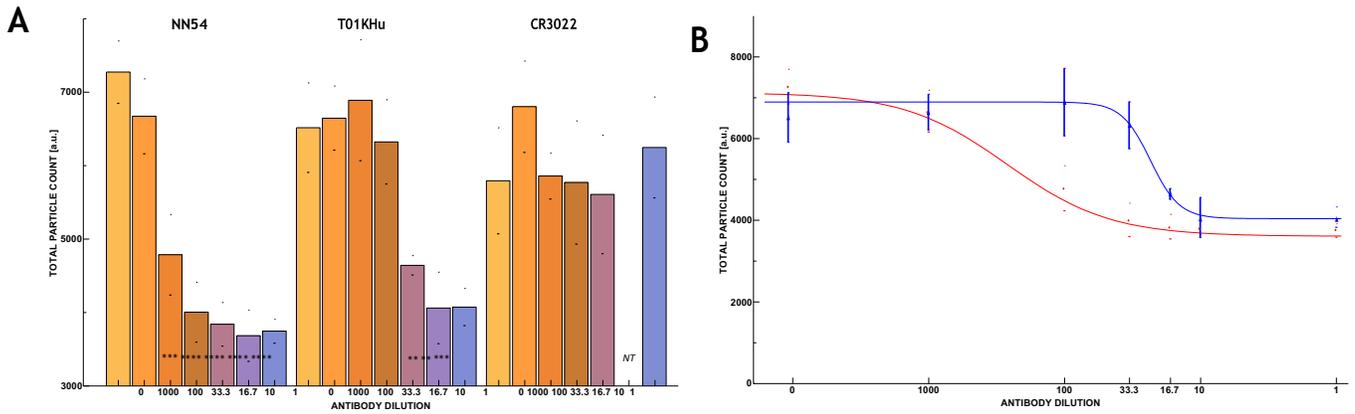

**Figure 3.** A) Neutralizing activity of neutralizing antibody clones NN54 (left) and T01KHu (middle), against non-neutralizing but binding antibody clone CR3022 (right), presented as total particle count of 10 images averaged over 3 repetitions for each reciprocal dilution factor. The highest concentration was 3.23 µg/ml. Each bar is tagged with the equivalent p-value star designation. NT means not tested. B) 4-parameter logistic curve fitted to neutralizing antibodies NN54 (red) and T01KHu (blue) data. $IC_{50}$ for NN54 and T01KHu were 12 ng/ml (80 pM, 1:269 dilution factor) and 138 ng/ml (933 pM, 1:23 dilution faction), respectively. The Hill coefficients for NN54 and T01KHu were calculated to be 1.148 and 4.0836, respectively, as described in section 2.4.

| Final antibody concentration (in µg/ml) | UCPV stock solution (0.4 µg/ml) volume used (in µl) | Antibody stock solution concentration (in µg/ml) | Dilution ratio | Antibody stock solution volume used (in µl) |
|---|---|---|---|---|
| 0.00323 | 300 | 0.1 | 1:1000 | 10 |
| 0.0323 | 300 | 1 | 1:100 | 10 |
| 0.0968 | 300 | 3 | 1:33.3 | 10 |
| 0.194 | 300 | 6 | 1:16.7 | 10 |
| 0.323 | 300 | 10 | 1:10 | 10 |
| 3.23 | 300 | 100 | 1:1 | 10 |

**Table 1.** Antibody concentrations and volumes used to prepare each final concentration.

# Supplementary Materials: Quantum Optical Immunoassay: Upconversion Nanoparticle-based Neutralizing Assay for COVID-19

**Navid Rajil**[1,+,×,†,‡], **Shahriar Esmaeili**[1,+,†,‡], **Benjamin W. Neuman**[1,2,3,+,‡], **Reed Nessler**[1,+,‡], **Hung-Jen Wu**[4,+,‡], **Zhenhuan Yi**[1,+,‡], **Robert W. Brick**[1,+,‡], **Alexei V. Sokolov**[1,5,+,‡], **Philip R. Hemmer**[1,6,7,+,‡], **and Marlan O. Scully**[1,5,*,+,‡]

[1]Institute for Quantum Science and Engineering, Texas A&M university, TX 77843, US
[2]Department of Biology, Texas A&M University, College Station, TX 77843, US
[3]Global Health Research Complex, Texas A&M University, College Station, TX 77843, US
[4]Department of Chemical Engineering, Texas A&M University, College Station, TX 77843, US
[5]Baylor University, Waco, TX 76798, US
[6]Department of Electrical & Computer Engineering, Texas A&M University, College Station, TX 77843, US
[7]Zavoisky Physical-Technical Institute, Federal Research Center "Kazan Scientific Center of RAS", Sibirsky Tract, 420029 Kazan, RU
[*]corresponding.scully@tamu.edu
[+]These authors conceptualized the experiment.
[×]This author built the microscope, prepared samples, acquired data.
[†]These authors processed the data.
[‡]These authors wrote the manuscript.

## ABSTRACT

In a viral pandemic, a few important tests are required for successful containment of the virus and reduction in severity of the infection. Among those tests, a test for the neutralizing ability of an antibody is crucial for assessment of population immunity gained through vaccination, and to test therapeutic value of antibodies made to counter the infections. Here, we report a sensitive technique to detect the relative neutralizing strength of various antibodies against the SARS-CoV-2 virus. We used bright, photostable, background-free, fluorescent upconversion nanoparticles conjugated with SARS-CoV-2 receptor binding domain as a phantom virion. A glass bottom plate coated with angiotensin-converting enzyme 2 (ACE-2) protein imitates the target cells. When no neutralizing IgG antibody was present in the sample, the particles would bind to the ACE-2 with high affinity. In contrast, a neutralizing antibody can prevent particle attachment to the ACE-2-coated substrate. A prototype system consisting of a custom-made confocal microscope was used to quantify particle attachment to the substrate. The sensitivity of this assay can reach 4.0 ng/ml and the dynamic range is from 1.0 ng/ml to 3.2 µg/ml. This is to be compared to 19 ng/ml sensitivity of commercially available kits.

## S1 Table of materials used

Table 1 shows the list of materials, company they were purchased from, and the part numbers.

| Materials | Company and catalog number |
|---|---|
| Streptavidin coated UCNP | Creative diagnostics # DNLC041 |
| Biotinylated receptor binding domain (RBD) | Acrobiosystems # SPD-C82E9 |
| Bovine serum albumin (BSA) | Sigma-Aldrich # A7030 |
| Tween 20 | Sigma-Aldrich |
| 10× PBS stock solution | Sigma-Aldrich # P5493-1L |
| Dopamine hydrochloride | Sigma-Aldrich # H8502 |
| Tris HCl | Thermofisher #15568025 |
| Nunc Labtek II 8-well bottom cover glass plates | Thermofisher # 155409 |
| goat anti-mouse IgG with Alexa Fluor 633 | Thermofisher # A-21052 |
| Angiotensin-converting enzyme 2 (ACE-2) | Raybiotech # 230-30165 |
| Recombinant SARS-CoV-2 RBD with C-terminal mouse IgG Fc Tag | Raybiotech # 230-30166 |
| Mouse anti-SARS-CoV-2 neutralizing antibody clone NN54 | Creative diagnostics # CABT-CS064 |
| Human anti-SARS-CoV-2 neutralizing antibody clone T01KHu | Thermofisher # 703958 |
| Human anti-SARS-CoV-2 non-neutralizing antibody but binding clone CR3022 | Absolute antibodies # AB01680-10.0 |

**Table S1** – Materials used in this work

## S2 Experimental methods

### S2.1 upconverting particles Phantom virion (UCPV) preparation
$NaYF_4,Yb,Er@NaYF_4$ upconversion nanoparticles (UCNPs) coated with streptavidin were purchased from Creative Diagnostics. The biotinylated RBD was purchased from Acrobiosystem. UCNPs coated with biotinylated RBD via streptavidin and biotin binding serves as phantom viruses. To prepare the phantom virus UCNPs, a 200 μl solution 0.5 mg/ml of UCNPs was mixed with 10 μl of 0.2 mg/ml biotinylated RBD and incubated for 1 hour on a shaker. Then the particles were washed 3 times (as described below) and resuspended in assay wash buffer (1×PBS, 0.5 % BSA, 0.1 % tween-20). After resuspension, a 0.4 μg/ml solution was prepared and sterile filtered and kept at 4 C until use.

### S2.2 Particle wash protocol
To wash the phantom virus particles after conjugation of biotinylated RBD on to streptavidin coated UCNPs, the particles were centrifuged at 9000 g for 10 minutes. 180 μl of supernatant was removed and replaced with 180 μl of assay wash buffer (1×PBS, 0.5% BSA, 0.1% tween 20). Then the particles were resuspended and sonicated in bath sonicator at 60 W power for 10 minutes. The bath water was constantly changed every 2-3 minutes with fresh ice-cold water to keep the phantom virus particles cold. This process was repeated 3 times with one exception. For the last wash, after removing 180 μl of supernatant, only 170 μl of assay wash buffer was added to raise the phantom particles final volume to 200 μl (concentration 0.5 mg/ml). Then a volume of 5 ml of 0.4 μg/ml phantom particle solution was made and filtered using 0.2 μm cellulose acetate syringe filters. One must note that 0.5 to 1 ml of final solution will be lost in filtration step and this amount must be considered to prevent shortage of particle solution.

### S2.3 ACE-2/polydopamine coating of the glass plates
ACE-2 proteins were coated onto glass substrates using the published polydopamine modification protocol1. Briefly, a 10 mM solution of Tris-HCl solution (PH = 8) was prepared and used to prepare a 2 mg/ml solution of dopamine hydrochloride. The solution then was mixed with 0.75 mg/ml ACE-2 protein solution at 1:1 volume ratio. The mixture was then plated on treated Nunc Labtek II 8-well bottom cover glass plates at 10 μl per well. The plates were incubated for 2 hours at room temperature in a humidity chamber to prevent drying. After 2 hours of incubation, each well was washed with 500 μl wash buffer (1×PBS, 0.5% BSA, 0.1%Tween 20) 4 times and incubated with 500 μl per well of blocking buffer (1×PBS, 5% BSA, 0.1%Tween 20) for 1 hour. After blocking, each well was washed with 500 μl of washing buffer 4 times. The plates were freshly prepared prior to use.

### S2.4 Examination of ACE-2/polydopamine coating of the glass plates
To examine ACE-2/polydopamine coating, we used RBD with mouse IgG Fc tag (RBD–FC) to identify ACE-2. Briefly, the prepared plates were incubated with 250 μl of RBD–Fc at a concentration of 10 μg/ml in washing buffer for 1 hour. Then, we washed the plates 4 times with washing buffer before adding 250 μl of the secondary antibody goat anti-mouse IgG with Alexa Fluor 633 at concentration of 10 μg/ml to detect the RBD. For comparison, negative control plates without RBD–FC or secondary antibodies were prepared. Samples without blocking were evaluated as well. The assay structure of this experiment is shown in figure 1 (a-f). The interactions between UCPVs and the polydopamine/ACE-2 coated plates were evaluated as well. For these test, we prepared a solution of 10 μg/ml UCPV and coated a prepared plate with 290 μl of this solution. To test the nonspecific binding between polydopamine and UCPVs, we prepared another plate coated with only polydopamine (mixed with 1× PBS instead of ACE-2 protein at 1:1 ratio). After 1 hour of incubation at room temperature, we washed the plates 4 times with wash buffer, and the plates were air dried at room temperature. Then, the fluorescent detection was conducted with the confocal microscope.

**S2.5 Upconversion nanoparticle-based antibody neutralization assay (UNIK)**

After preparation of UCPV, a dilution of 1 μg/ml was prepared. Seven vials of 300 μl of 0.4 μg/ml particles were separated and 10 μl of different dilutions of antibody solution in wash buffer were added to each vial such that each vial received only one dilution of antibody sample. The samples were incubated on a shaker for 1 hour. This step was done in parallel to blocking the step of the plate preparation. After incubation of UCPV and antibody and the blocking of the plates, the plates were washed 4 times and 300 μl of each UCPV sample was added to separate wells of the plates and incubated for 1 hour on tilt shaker. After this incubation, the wells were washed 5 times with wash buffer and imaged for particle count. The plates were stored at 4 C until imaging.

**S2.6 Data acquisition and processing**

*S2.6.1 ACE-2 protein coating examination data acquisition and processing*

As described in section S2.4, we prepared several samples to check the coating of ACE-2 protein on the glass coverslip plates. As described, the ACE-2-coated plates were coated with RBD tagged with mouse IgG Fc. The RBD was detected using goat anti mouse IgG conjugated with Alexa Fluor 633. To detect the fluorescence, a custom-made laser scanning confocal microscope was used. The schematic of the microscope is shown in figure S1 (supplementary materials). From each sample, spectra of 25 points from a 5 by 5 grid were collected and averaged. The excitation laser was a 638 nm laser, and the laser's output power was set to 10 mW for all measurements. The acquisition time was 1 second. The collected spectrum for each negative and positive control sample was averaged and plotted. The results are shown in Figure S4 b, d, f, and h.

*S2.6.2 UNIK data acquisition and processing*

Data acquisition for up-conversion based antibody neutralization kit (described in section S2.5) and UCPV specific and nonspecific binding to polydopamine/ACE-2 and polydopamine coatings respectively (section S2.4) were done with some differences relative to ACE-2 coating examination (section S2.4).

A multimode high-power laser was focused on each sample with a 50 μm diameter spot size using an oil immersion objective (Leica HCX Plan Apo 40×/1.25-0.75 OIL CS ¥/0.17/E objective). The input power to the objective was measured to be 300 mW (Figure S2, measured at point A). Each data point was scanned 10 times. Each scan was 145 μm by 145 μm and this area was imaged using a raster scan of 8 × 8 points. The fluorescent image from the particles was then reflected onto an ICCD camera (Starlight Xpress Trius Pro 674). The effective imaged area was 87 μm by 145 μm for the antibody titer tests. Each point of the 8 by 8 raster scan was integrated for 200 ms, resulting in acquisition time of 12.8 seconds per image. Each sample was imaged 10 times on a 2 by 5 grid. The step size for this grid was 500 μm. The scanning, imaging, and optical setup details can be found in Figure S2. Image data were saved and reconstituted in Mathematica using a custom-made code. The software was used to count UCPV foci in 10 fields of view per test per data point, and then summed and averaged over 3 repetitions to yield the particle counts. Thus, the error bars in Figure 3a and 3b in the main article show the fluctuation of number of particles counted for each data point across 3 repetitions

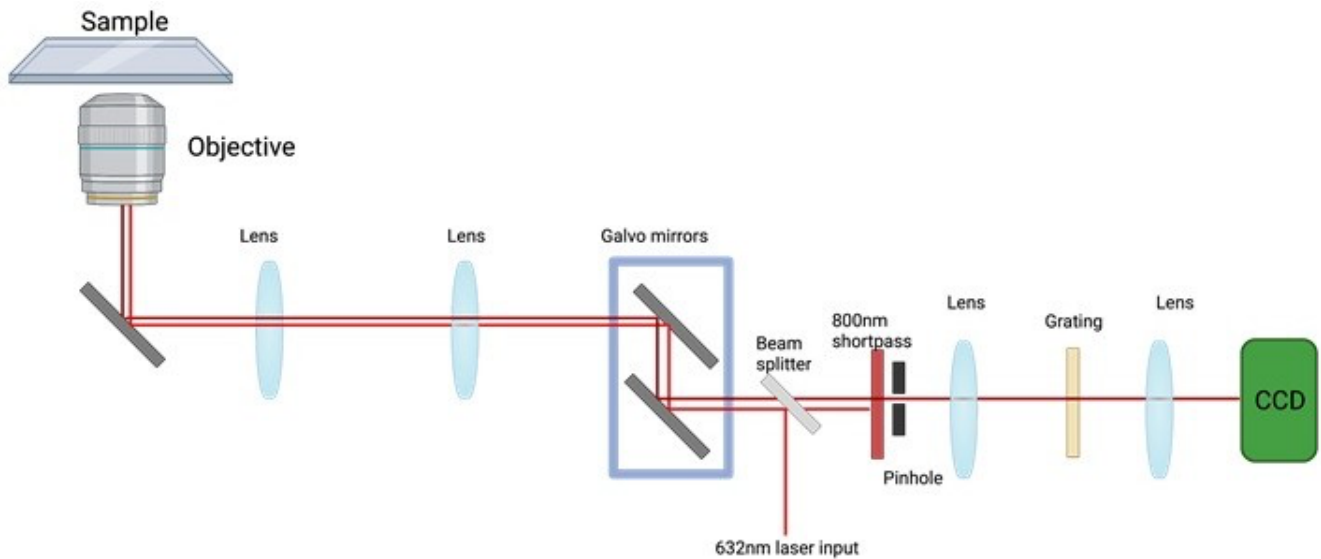

**Figure S1** – Confocal spectroscopy and microscopy setup schematic. Using the galvo mirrors, the 632 nm laser is moved on the sample in a raster scan. A 632 nm notch filter (shown in red rectangle before pinhole) blocks the laser residual reflection. The image of the pinhole is blown through a transmission grating 300 lines/mm and imaged on an ICCD camera. The pixel counts are then transformed to spectrum data.

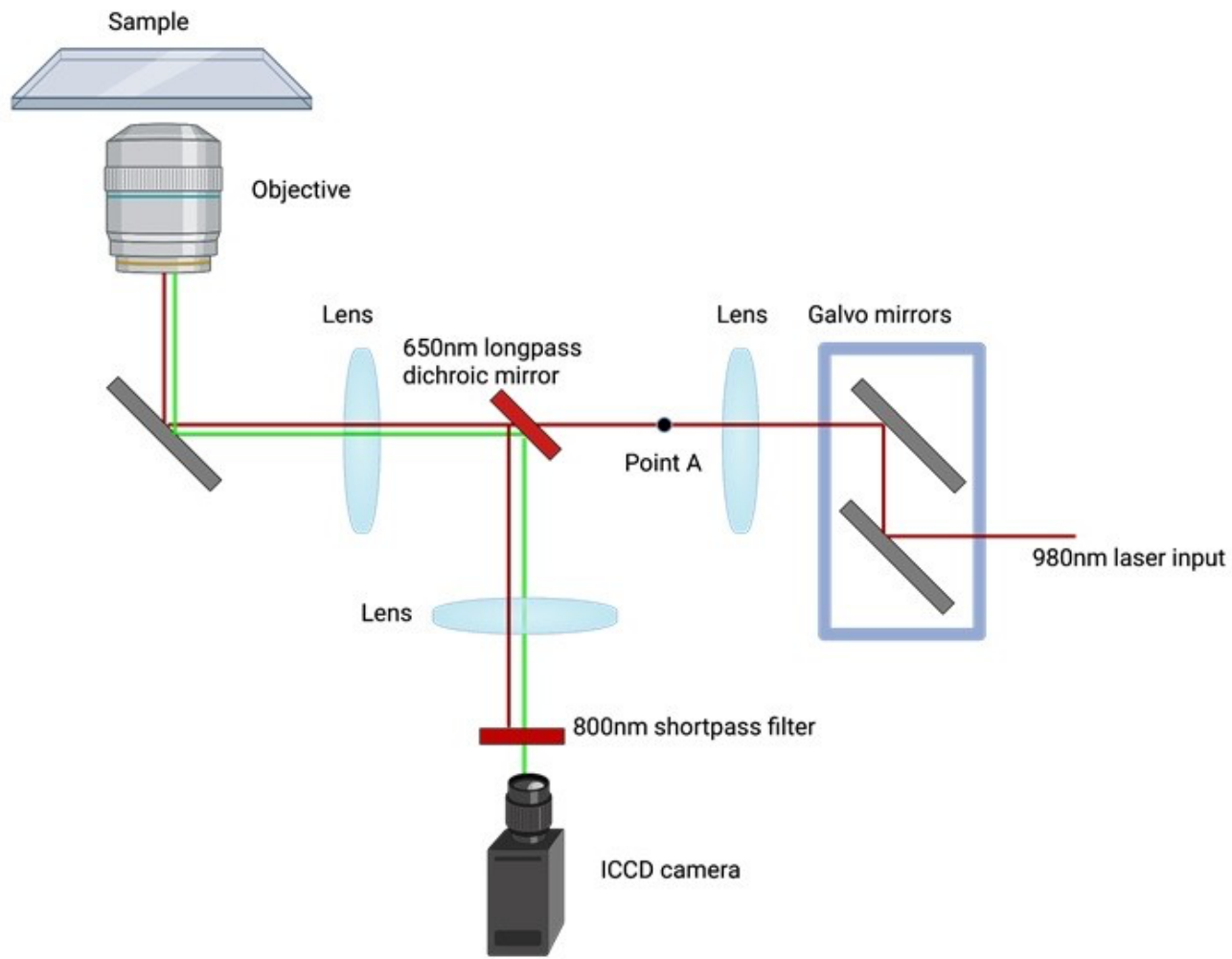

**Figure S2** – Wide field fluorescence imaging setup schematic. Using galvo mirrors, the laser is moved on the sample in a raster scan. The ICCD camera collects the signal from each point of the raster scan during the whole time of the scan.

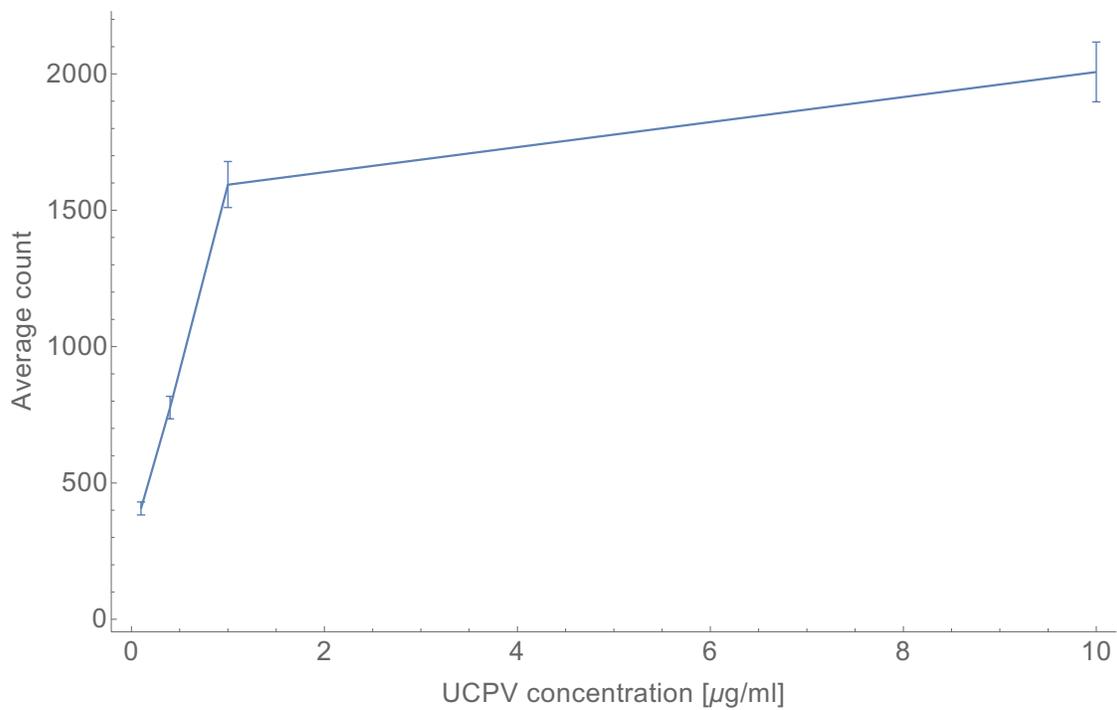

**Figure S3** – Optimization of UCPV concentration. Average count per image of 5 images versus concentrations of UCPV at 0.1 µg/ml, 0.4 µg/ml, 1 µg/ml, and 10 µg/ml is plotted to choose optimized concentration of UCPV.

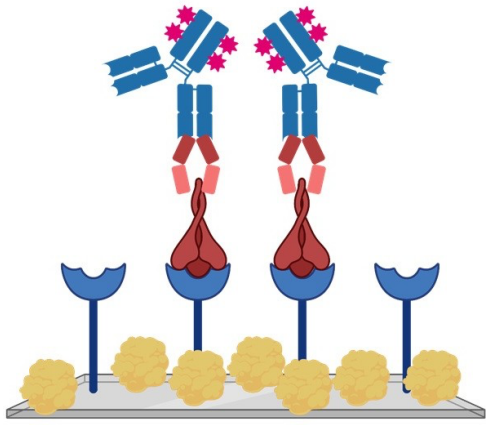 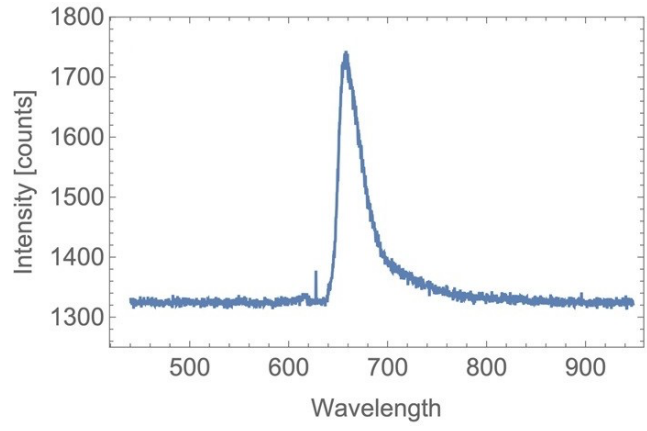

(a)

(b)

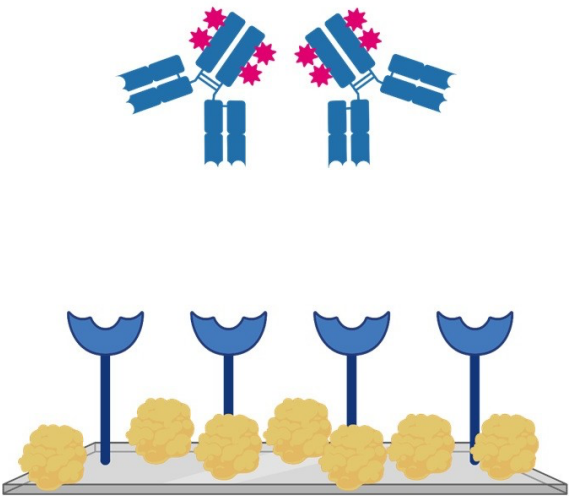 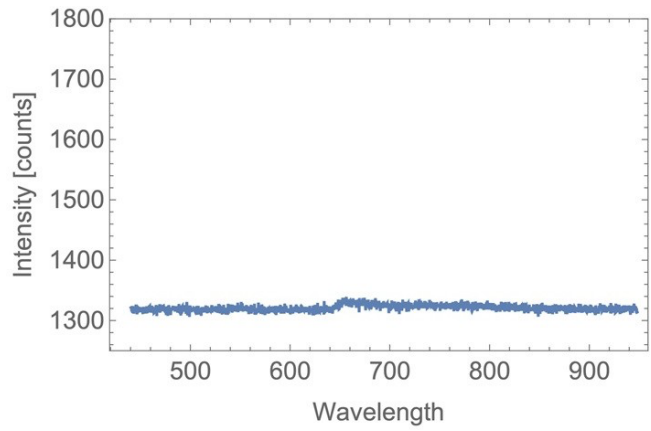

(c)

(d)

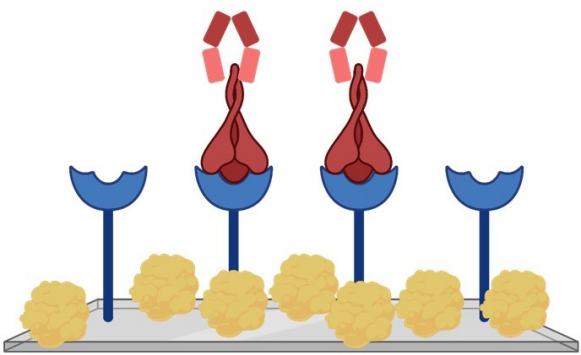 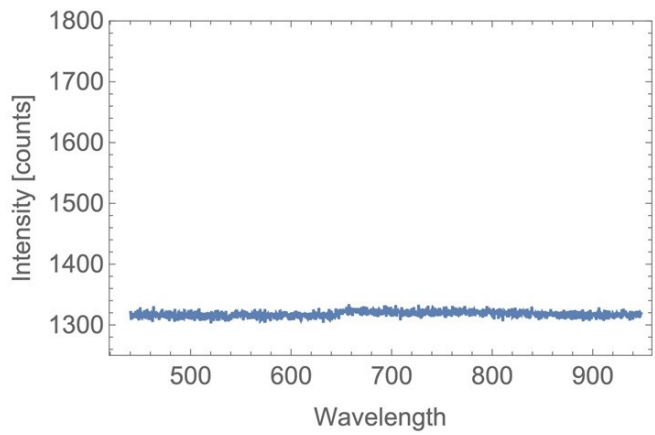

(e)

(f)

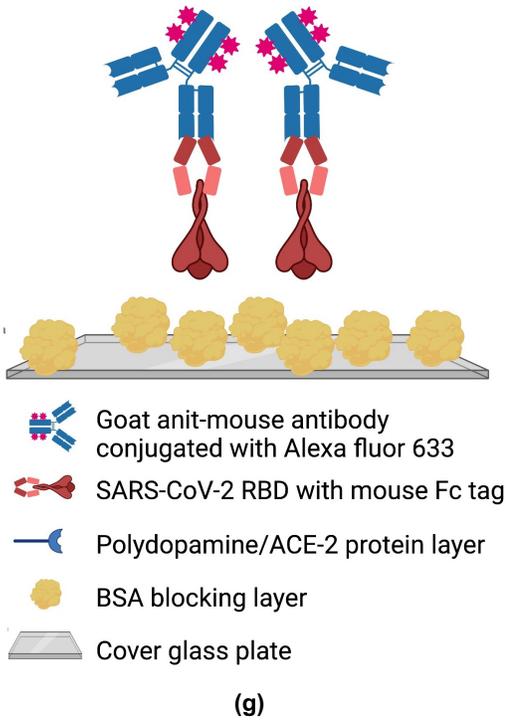
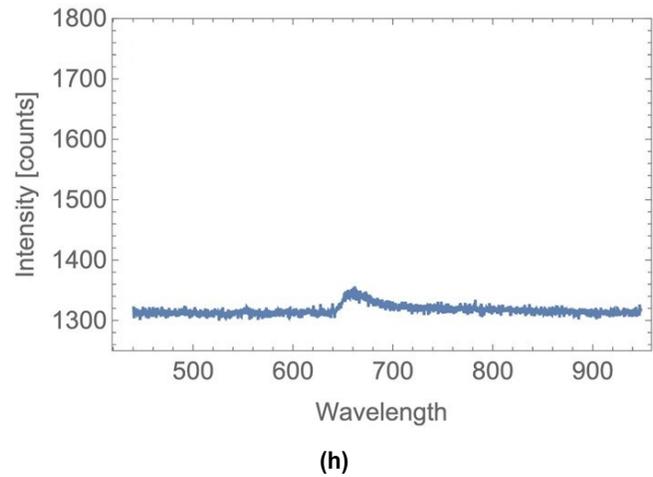

(g)

(h)

**Figure S4** – ACE-2 coating test assay structure (a, c, e, g) and corresponding spectrum results (b, d, f, h). a) Full positive assay including ploydopamine/ACE-2 coating, blocking, RBD–FC, and the secondary antibody. c) Full negative control assay Missing RBD–FC. e) Positive assay missing secondary antibody to assess the autofluorescence background. g) Positive assay Missing ACE-2 protein (coated with polydopamine mixed with 1×PBS) to assess the non-specific background. b) Full positive assay including ploydopamine/ACE-2 coating, blocking, RBD–FC, and secondary antibody. A strong fluorescence from secondary antibody is observed. d) Full negative control assay Missing RBD–FC. A zero-fluorescence signal is expected and observed. f) Positive assay Missing secondary antibody to assess the autofluorescence background. No auto fluorescence background is observed. h) Positive assay Missing ACE-2 protein (coated with polydopamine mixed with 1× PBS) to assess the non-specific background in which a very weak signal was observed. Since 2c shows no fluorescence, considering assay structure, this signal most probably is due to nonspecific binding between RBD–FC and the plate.

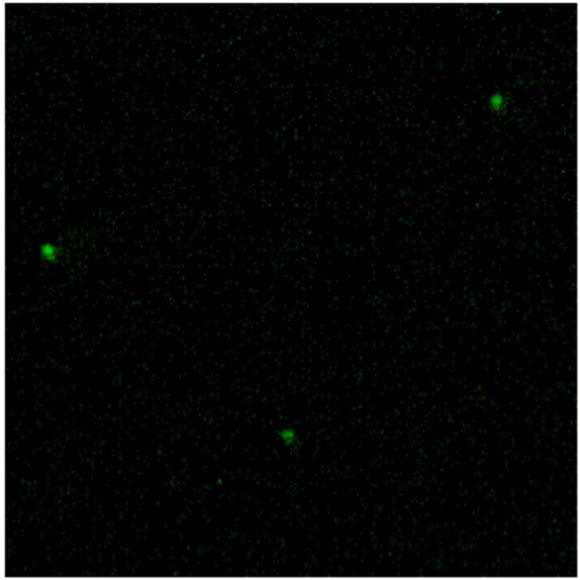 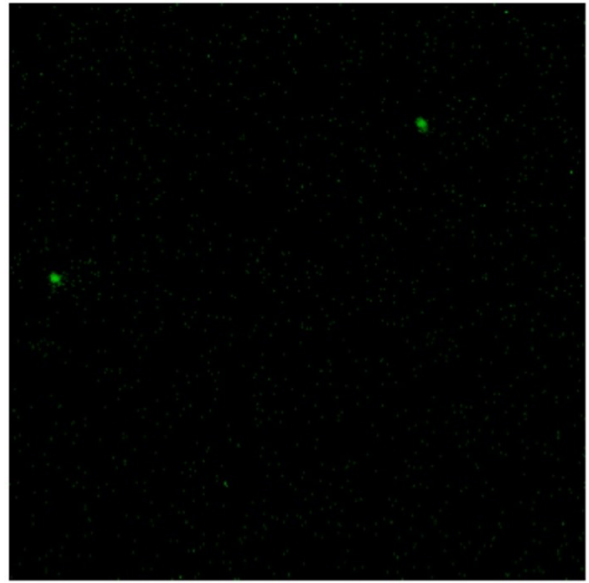

(a) (b)

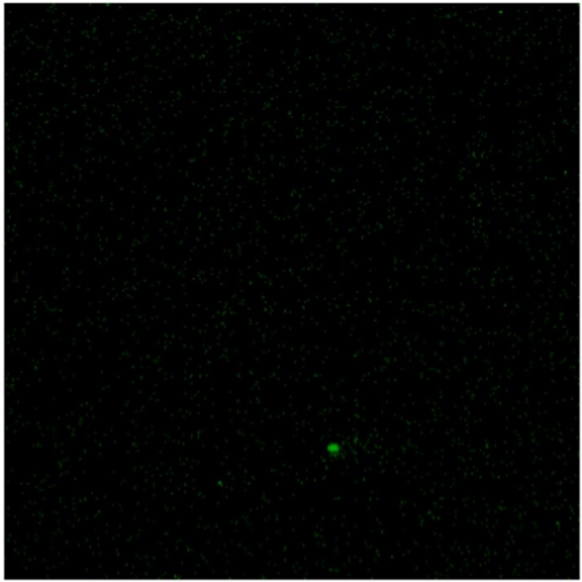

(c)

**Figure S5** – Affinity of UCPV and Polydopamine/PBS coated area. a, b, c) shows 3 scans of 3 different areas in the center of circular area coated with polydopamine/PBS to assess the nonspecific binding between UCPVs and blocked polydopamine. Small green spots are visible in each figure and represent a particle. This shows that the binding shown in Figure 2b is due to the intrinsic affinity between ACE-2 protein and RBD. All images are 53 $\mu$m by 53 $\mu$m.

## S3 4-parameter logistic function fits

The 4-parameter logistic curves were fitted using the online tool available at ATT Bioquest[2]. For neutralizing antibody type 1 (NN54), the equation, IC50 (midpoint), and Hill coefficient are below:

$$Y_{NN54} = B + \frac{A - B}{1 + (con_{NN54}/IC50)^{hc}} \tag{S1}$$

where $A=7123$, $B=3603$, $IC50=0.0122$, $hc=1.148$, and $con_{NN54}$ is the concentration of antibody (clone NN54). So:

$$IC50 = 0.0122 \; \mu g/ml \text{ and Hill coeff.} = 1.148 \tag{S2}$$

For neutralizing antibody type 2 (T01KHu), the equation, IC50 (midpoint), and Hill coefficient are shown below:

$$Y_{T01KHu} = B + \frac{A-B}{1+(con_{T01KHu}/IC50)^{hc}} \tag{S3}$$

where $A=6892$, $B=4035$, $IC50=0.138$, and $hc=4.084$, and $con_{T01KHu}$ is the concentration of antibody (clone T01KHu). So,

$$IC50 = 0.138 \; \mu g/ml \text{ and Hill coeff.} = 4.084 \tag{S4}$$

## S4 Limit of detection (LOD) calculation

The standard deviation for NN54 negative control sample was 423 counts. Thus, using S1:

$$x_{LOD}^{NN54} = 0.004 \; \mu g/ml \tag{S5}$$

The standard deviation for T01KHu negative control sample was 608 counts. Thus, using S3:

$$x_{LOD}^{T01KHu} = 0.128 \; \mu g/ml \tag{S6}$$

## S5 Calculated *p*-values for each data point

To calculate the P value, we performed the T-test using the built-in function in Mathematica software. Total counts of each data point (set of 3 numbers from the 3 repetitions) were compared with the set of total counts negative control (data point with no antibody) data set.

| Antibody type | *p*-value 0.00323 µg/ml | *p*-value 0.0323 µg/ml | *p*-value 0.0968 µg/ml | *p*-value 0.194 µg/ml | *p*-value 0.323 µg/ml | *p*-value 3.23 µg/ml |
|---|---|---|---|---|---|---|
| NN54 (neutralizing type1) | NS | 3.4x10⁻³ | 6.5x10⁻⁴ | 3.3x10⁻⁴ | 3.5x10⁻⁴ | 1.8x10⁻⁴ |
| T01KHu (neutralizing type2) | NS | NS | NS | 6.5x10⁻³ | 5.5x10⁻³ | 3.0x10⁻³ |
| CR3022 (Non-neutralizing type3) | NS | NS | NS | NS | NT | NS |

**Table S2** – *p*-value of total counts of 3 repetitions of each data point for each antibody calculated against the set of 3 repetition of negative control counts. One star when *p*-value ≤ 0.05, two stars when *p*-value ≤ 0.01, three stars when *p*-value ≤ 0.005, and four stars when *p*-value ≤ 0.001. NT stands for not tested. NS stands for non-significant.

## S6 KD value and θ approximation

This difference in IC50 also provides evidence for the inherent assay sensitivity which is limited by the affinity between antibody and antigen[3]. For instance, in the equation below for protein *P* binding with ligand *L* and producing protein–ligand complex *PL*:

$$P + L \rightleftharpoons PL \tag{S7}$$

The dynamic equation for the concentration of protein–ligand complex [PL] can be written as

$$\frac{d[PL]}{dt} = \kappa_a [P]_f [L]_f - \kappa_d [PL] \tag{S8}$$

Where $\kappa_a$ is the association rate of protein and ligand, $\kappa_d$ is the dissociation rate of protein–ligand complex, $[P]_f$ and $[L]_f$ are free protein and free ligand concentration. At equilibrium this equation is equal to zero $\frac{d[PL]}{dt} = \kappa_a [P]_f [L]_f - \kappa_d [PL]_f = 0$, which leads to:

$$\frac{[P]_f [L]_f}{[PL]} = \frac{\kappa_d}{\kappa_a} \equiv K_d. \tag{S9}$$

In a simple case, we can define the ratio of proteins-ligand concentration to total protein as:

$$\theta = \frac{[PL]}{[P]_t} \tag{S10}$$

with

$$[P]_f = [P]_t - [PL] \tag{S11}$$

where $[P]_t$ is the total concentration of protein. Substituting S11 in S9 and approximate $[L]_f = [L]_t$ (leading to the higher bound for θ) and rearranging will yield

$$\frac{([P]_t - [PL])[L]_t}{[PL]} \approx K_d \tag{S12}$$

Simple algebra and rearranging will yield

$$\theta = \frac{[PL]}{[P]_t} \approx \frac{[L]_t}{[L]_t + K_d} \tag{S13}$$

The parameter θ is the ratio of filled proteins and is related to $K_d$. θ = 0.5 when $[L]_t = K_d$. So, in some sense $K_d$ value (reported in Molar) is the concentration of ligand at which 50% of the proteins are filled with ligands. Now, since one basicall measures the number of filled proteins on the substrate in an ELISA-based assay (through any means of measurement), then it is easy to see that (as an approximation):

$$\theta = \frac{[PL]}{[P]_t} \approx \frac{[L]_t}{[L]_t + K_d} = 0.5 \; and \; is \; proportional \; to IC50 \tag{S14}$$

Thus, one can correlate IC50 point to the $K_d$ value.
As such, the best strategy to improve the limit of detection (LOD) for a certain antigen is to first acquire the best antibody with lowest dissociation constant ($K_d$ value) and implement the detection apparatus capable of achieving said sensitivity. In general, since $K_d$ value of antibodies for their targets varies between $10^{-5}$ M to $10^{-12}$ M from antibody to antibody, the detection assay's LOD will vary from antibody to antibody.

## S7 Exact solution for θ

$$\frac{[P]_f[L]_f}{[PL]} = K_d \tag{S15}$$

$$[P]_f + [PL] = [P]_t \tag{S16}$$

$$[L]_f + [PL] = [L]_t \tag{S17}$$

Using S16 and S17 in S15 and simple algebra will yield

$$[P]_t[L]_t - ([P]_t + [L]_t + K_d)[PL] + [PL]^2 = 0 \tag{S18}$$

Solving S18 for [PL] gets

$$[PL]_1 = \frac{([P]_t + [L]_t + K_d) - \sqrt{([P]_t + [L]_t + K_d)^2 - 4[P]_t[L]_t}}{2} \tag{S19}$$

$$[PL]_2 = \frac{([P]_t + [L]_t + K_d) + \sqrt{([P]_t + [L]_t + K_d)^2 - 4[P]_t[L]_t}}{2} \tag{S20}$$

Assuming a constant concentration for $[P]_t$ and only titer $[L]_t$, the second solution is unphysical since $\lim_{[L]_t \to \infty}[PL]_2 = \infty$ is not bounded. Thus only $[PL]_1$ is an acceptable solution. Thus:

$$\theta = \frac{[PL]}{[P]_t} = \frac{([P]_t + [L]_t + K_d) - \sqrt{([P]_t + [L]_t + K_d)^2 - 4[P]_t[L]_t}}{2[P]_t} \tag{S21}$$

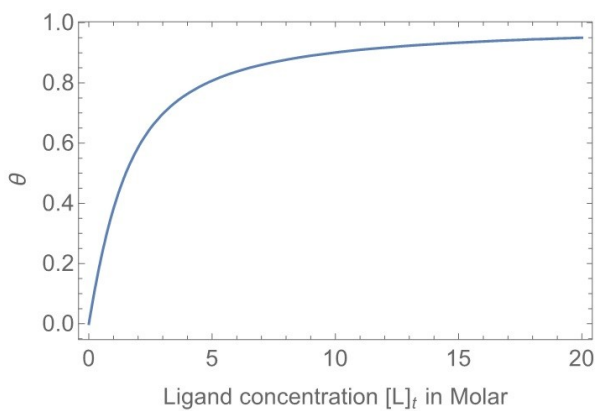 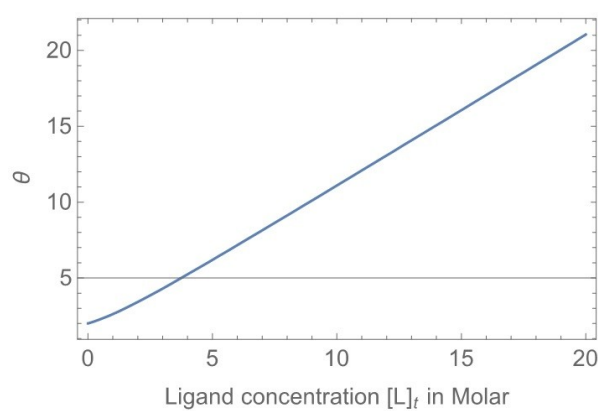

**Figure S6** – a) Plot of $[PL]_1$ for the case of $K_d = [P]_t = 1$. b) Plot of $[PL]_2$ for the case of $K_d = [P]_t = 1$.

## S8 The Theoretical solution for UNIK

The counted particles in the image are those that have bound to the ACE-2 protein on the substrate. Thus, we can write $\theta_1$ for this binding as

$$\theta_1 = \frac{[ACE.UCPV]}{[ACE]_t} = \frac{([ACE]_t + [UCPV]_f + K_d^{(1)}) - \sqrt{([ACE]_t + [UCPV]_f + K_d^{(1)})^2 - 4[ACE]_t[UCPV]_f}}{2[ACE]_t} \tag{S22}$$

It should be noted that the $[UCPV]_f$ here is the total available UCPVs (or UCPVs) that are not blocked by the antibodies. Thus, we must find the concentration of non-blocked RBDs. We can find the ratio of blocked RBDs using S24.

$$\theta_2 = \frac{[UCPV.Ab]}{[UCPV]_{total}} = \frac{([UCPV]_{total} + [Ab]_t + K_d^{(2)}) - \sqrt{([UCPV]_{total} + [Ab]_t + K_d^{(2)})^2 - 4[UCPV]_{total}[Ab]_t}}{2[UCPV]_{total}} \quad (S23)$$

Now, the total concentration of unblocked RBDs is $1 - \theta_2$

$$1 - \theta_2 = \frac{[UCPV]_f}{[UCPV]_{total}} = \quad (S24)$$
$$(([UCPV]_{total} - [Ab]_t - K_d^{(2)}) + \sqrt{([UCPV]_{total} + [Ab]_t + K_d^{(2)})^2 - 4[UCPV]_{total}[Ab]_t})/(2[UCPV]_{total})$$

Then

$$\frac{[UCPV]_f}{[UCPV]_{total}} = \quad (S25)$$
$$(([UCPV]_{total} - [Ab]_t - K_d^{(2)}) + \sqrt{([UCPV]_{total} + [Ab]_t + K_d^{(2)})^2 - 4[UCPV]_{total}[Ab]_t})/(2[UCPV]_{total})$$

$$[UCPV]_f = (([UCPV]_{total} - [Ab]_t - K_d^{(2)}) + \sqrt{([UCPV]_{total} + [Ab]_t + K_d^{(2)})^2 - 4[UCPV]_{total}[Ab]_t})/(2) \quad (S26)$$

This is the available RBD concentration that can bind to the substrate and be counted. Figure S6 shows the plot of $q_1$ when S27 is plugged in, as function of total antibody concentration $[Ab]_t$. The parameter for this plot were $[RBD]_{initial} = [ACE]_t = K_d^{(1)} = K_d^{(2)} = 1$. It is evident that the LOD depends on ACE-2 concentration, initial RBD (UCPV) concentration, and $K_d$ values of binding between RBD and antibody ($K_d^{(2)}$) and RBD and ACE-2 ($K_d^{(2)}$) protein. Only two can be controlled by assay developer, ACE-2 concentration and RBD (UCPV) concentration.

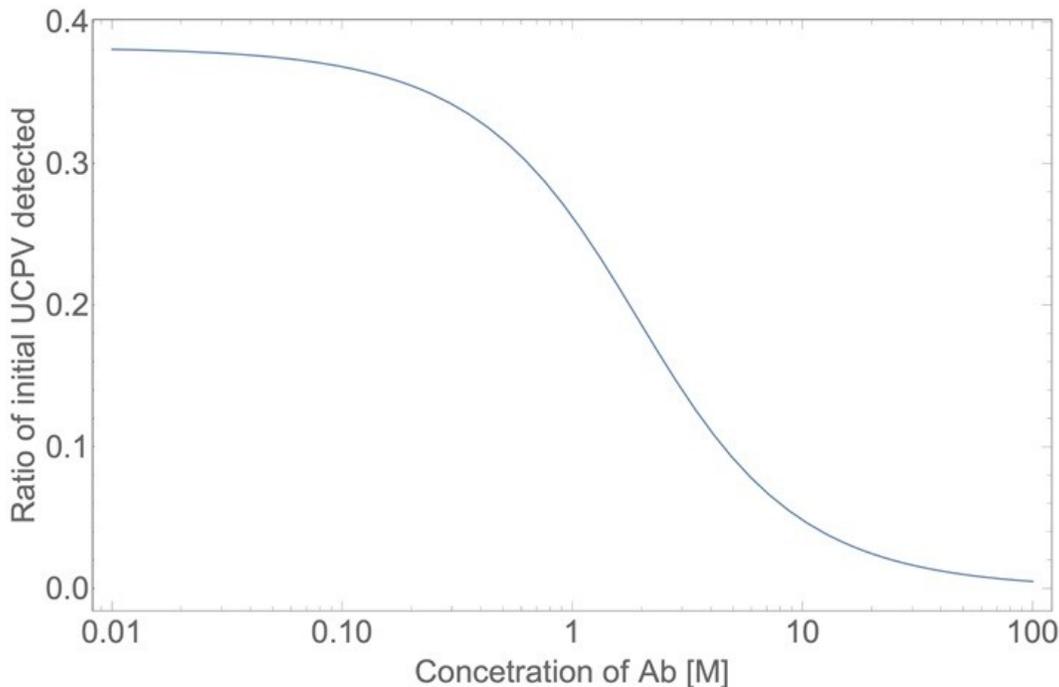

**Figure S7** – Theoretical calibration plot of UNIK when $[UCPV]_{total} = [ACE]_t \ [ACE]_t = K_d^{(1)} = K_d^{(2)} = 1$.

## S9 Derivation $[Ab]_t|_{IC50}$ for UNIK

From S22 we can find the $[UCPV]_f|_{IC50}$ when 50% of ACE protein on the substrate is full, with simple algebra as

$$[UCPV]_f|_{IC50} = \frac{1}{2}[ACE] + K_d^{(1)}. \tag{S27}$$

Plugging in the S26 we get (note that $[UCPV]_f|_{IC50}$ happens at a certain antibody concentration which we show it as $[Ab]_t|_{IC50}$)

$$\frac{([UCPV]_{total} - [Ab]_t|_{IC50} - K_d^{(2)}) + \sqrt{([UCPV]_{total} + [Ab]_t|_{IC50} + K_d^{(2)})^2 - 4[UCPV]_{total}[Ab]_t|_{IC50}}}{2} = \frac{1}{2}[ACE] + K_d^2. \tag{S28}$$

Solving for $[Ab]_t|_{IC50}$ will yield

$$[Ab]_t|_{IC50} = [UCPV]_{total} - K_d^{(2)} + \frac{2[UCPV]_{total}K_d^{(2)}}{[ACE] + 2K_d^{(1)}} - \frac{1}{2}[ACE] - K_d^{(1)}. \tag{S29}$$

There are two basic assumptions inherit in S33. One is $[UCPV]_{total} \neq 0$ and $[ACE]_{total} \neq 0$. The second is the fact that $[Ab]_t|_{IC50} > 0$. This second condition will give us the following constraint:

$$[UCPV]_{total} > \frac{1}{2}[ACE] + K_d^{(1)}. \tag{S30}$$

With $K_d^{(1)} = 10^{-9}$ M, and $K_d^{(2)} = 10^{-12}$ M we will get the following graph for the general behavior of $[Ab]_t|_{IC50}$ as a function of [ACE] for different $[UCPV]_{total}$ concentrations which shows that we must decrease UCPV concentration and maximize ACE-2 protein concentration

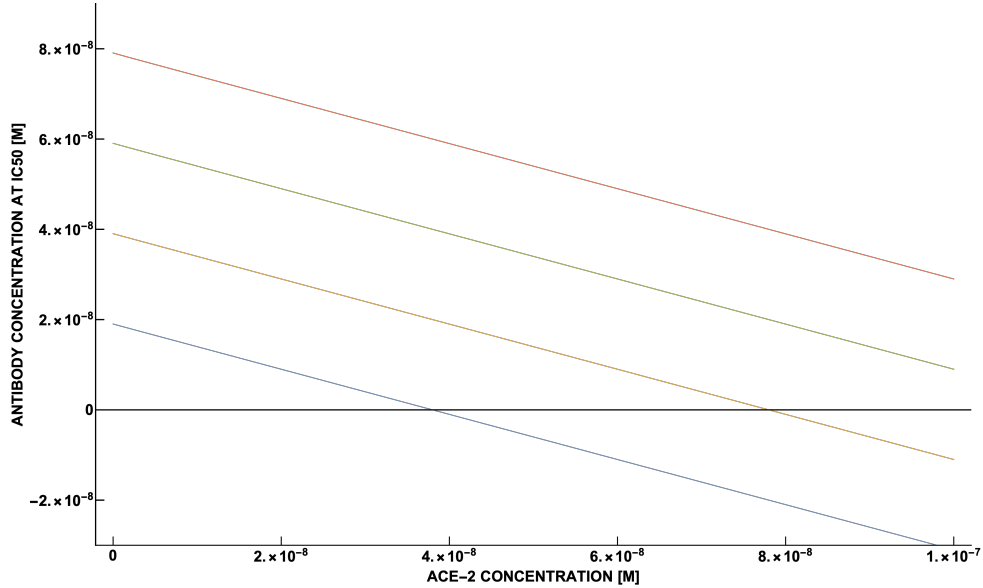

**Figure S8** – Behavior of $[Ab]_t|_{IC50}$ as a function of ACE-2 concentration for different concentrations of $[UCPV]_{total}$. Red, green, orange, and blue indicate $[UCPV]_{total} = 8 \times 10^{-8}$, $[UCPV]_{total} = 6 \times 10^{-8}$, $[UCPV]_{total} = 4 \times 10^{-8}$, $[UCPV]_{total} = 2 \times 10^{-8}$ molar concentrations respectively. Other constants were $K_d^{(1)} = 10^{-9}$ M and $K_d^{(2)} = 10^{-12}$ M.

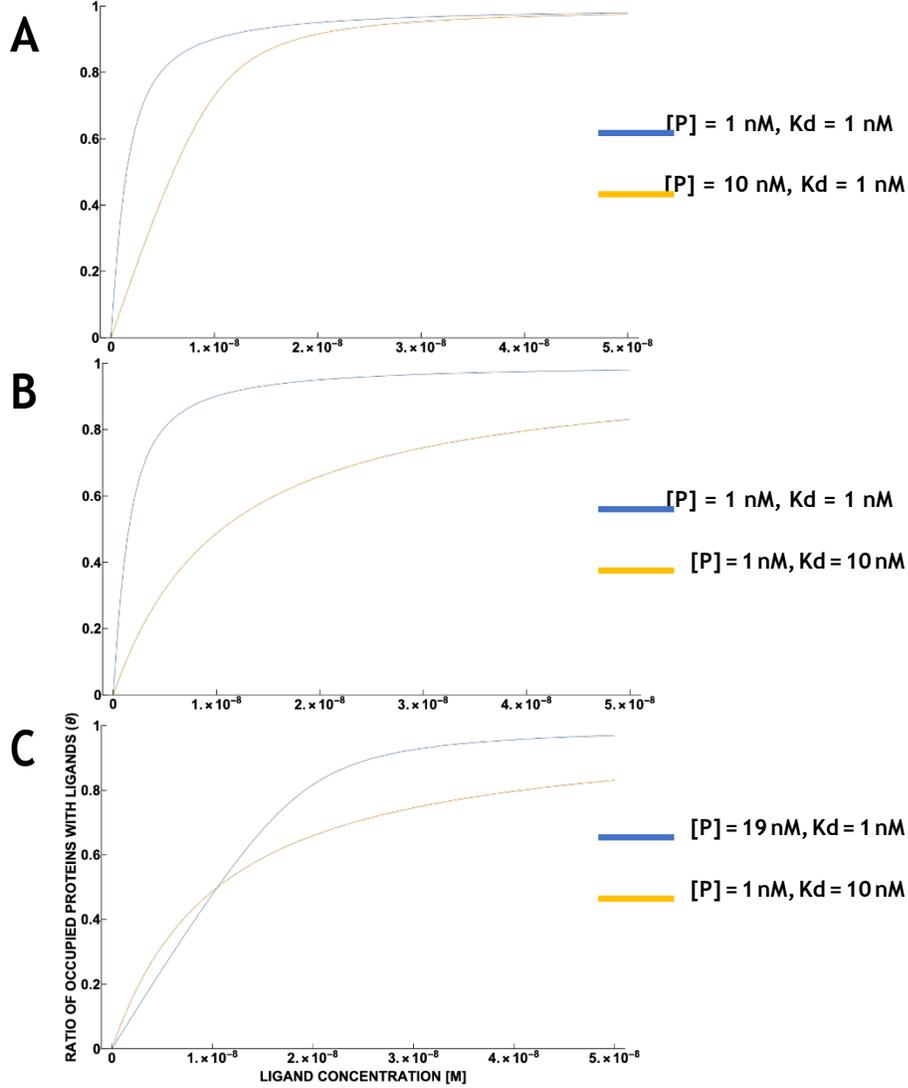

**Figure S9** – Numerical solutions of $\theta$ in Eq. 2 for A) $[P]_t = 1nM$, $K_d = 1nM$ vs $[P]_t = 10nM$, $K_d = 1nM$. B) $[P]_t = 1nM$, $K_d = 1$ vs $[P]_t = 1nM$, $K_d = 10nM$, and C) $[P]_t = 19nM$, $K_d = 1nM$ vs $[P]_t = 1nM$, $K_d = 10nM$.

### S10 A quantum description of $\theta$

We consider a bimolecular reaction $A + B \rightleftharpoons AB$ where the initial populations of A and B are $a$ and $b$, respectively. The population $n$ of AB is modeled probabilistically as a birth–death process[4], where the birth rate from state $n$ to $n+1$ is $\lambda_n = \alpha(a-n)(b-n)$ and the death rate from $n$ to $n-1$ is $\mu_n = \beta n$. Here $\alpha$ and $\beta$ are rate constants particular to the reaction. Using $p_n$ ($n = 0, 1, \ldots, \min(a,b)$) to denote the probability distribution of $n$, we have the governing equations

$$\frac{dp_0}{dt} = -\lambda_0 p_0 + \mu_1 p_1 \tag{S31}$$

$$\frac{dp_{0n}}{dt} = -\lambda_n p_n + \lambda_{n-1} p_{n-1} - \mu_n p_n + \mu_{n-1} p_{n-1} \text{ when } n \geq 1 \tag{S32}$$

whose steady-state ($\frac{dp_n}{dt} = 0$) solution is

$$p_n = p_0 \prod_{k=1}^{n} \lambda_{k-1}/\mu_k \tag{S33}$$

with $p_0$ determined by the normalization $\sum_n p_n = 1$. With the given birth and death rates this distribution is described by the probability-generating function

$$G(z) = \sum_{n=0}^{\min(a,b)} p_n z^n = \frac{{}_2F_0(-a,-b;;\frac{\alpha}{\beta}z)}{{}_2F_0(-a,-b;;\frac{\alpha}{\beta})} \tag{S34}$$

where ${}_2F_0$ is a generalized hypergeometric function. The expected population of AB at equilibrium is

$$G'(1) = ab\frac{\alpha}{\beta}\frac{({}_2F_0(1-a,1-b;;\frac{\alpha}{\beta}))}{{}_2F_0(-a,-b;;\frac{\alpha}{\beta})}. \tag{S35}$$

When evaluated numerically this expression is found to agree closely with the non-probabilistic treatment of the law of mass action, i.e. where the equilibrium value of $n$ solves the equation

$$((a-n)(b-n))/n = \beta/\alpha, \tag{S36}$$

which resembles Eqn. S15.